\documentclass{aastex631} 

\newcommand{\objname}{2010 LH$_{15}$}
\newcommand{\objnameFull}{2010 LH$_{15}$ (alternately designated 2010 TJ$_{175}$)} 

\begin{document}

\title{New Recurrently Active Main-belt Comet 2010 LH15} 

\correspondingauthor{Colin Orion Chandler}
\email{coc123@uw.edu}

\author[0000-0001-7335-1715]{Colin Orion Chandler}
\affiliation{Dept. of Astronomy \& the DiRAC Institute, University of Washington, 3910 15th Ave NE, Seattle, WA 98195, USA}
\affiliation{LSST Interdisciplinary Network for Collaboration and Computing, 933 N. Cherry Avenue, Tucson, AZ 85721, USA}
\affiliation{Dept. of Astronomy \& Planetary Science, Northern Arizona University, PO Box 6010, Flagstaff, AZ 86011, USA}

\author[0000-0001-5750-4953]{William J. Oldroyd}
\affiliation{Dept. of Astronomy \& Planetary Science, Northern Arizona University, PO Box 6010, Flagstaff, AZ 86011, USA}

\author[0000-0001-7225-9271]{Henry H. Hsieh}
\affiliation{Planetary Science Institute, 1700 East Fort Lowell Rd., Suite 106, Tucson, AZ 85719, USA}
\affiliation{Institute of Astronomy and Astrophysics, Academia Sinica, P.O.\ Box 23-141, Taipei 10617, Taiwan}

\author[0000-0001-9859-0894]{Chadwick A. Trujillo}
\affiliation{Dept. of Astronomy \& Planetary Science, Northern Arizona University, PO Box 6010, Flagstaff, AZ 86011, USA}

\author[0000-0002-6023-7291]{William A. Burris}
\affiliation{Dept. of Physics, San Diego State University, 5500 Campanile Drive, San Diego, CA 92182, USA}
\affiliation{Dept. of Astronomy \& Planetary Science, Northern Arizona University, PO Box 6010, Flagstaff, AZ 86011, USA}

\author[0000-0001-8531-038X]{Jay K. Kueny}
\altaffiliation{National Science Foundation Graduate Research Fellow}
\affiliation{University of Arizona Dept. of Astronomy and Steward Observatory, 933 North Cherry Avenue Rm. N204, Tucson, AZ 85721, USA}
\affiliation{Lowell Observatory, 1400 W Mars Hill Rd, Flagstaff, AZ 86001, USA}
\affiliation{Dept. of Astronomy \& Planetary Science, Northern Arizona University, PO Box 6010, Flagstaff, AZ 86011, USA}

\author[0000-0002-7489-5893]{Jarod A. DeSpain}
\affiliation{Dept. of Astronomy \& Planetary Science, Northern Arizona University, PO Box 6010, Flagstaff, AZ 86011, USA}

\author[0000-0003-2521-848X]{Kennedy A. Farrell}
\affiliation{Dept. of Astronomy \& Planetary Science, Northern Arizona University, PO Box 6010, Flagstaff, AZ 86011, USA}

\author[0000-0002-2204-6064]{Michele T. Mazzucato} 
\altaffiliation{Active Asteroids Citizen Scientist}
\affiliation{Royal Astronomical Society, Burlington House, Piccadilly, London, W1J 0BQ, UK}

\author[0000-0002-9766-2400]{Milton K. D. Bosch} 
\altaffiliation{Active Asteroids Citizen Scientist}

\author{Tiffany Shaw-Diaz} 
\altaffiliation{Active Asteroids Citizen Scientist}

\author{Virgilio Gonano} 
\altaffiliation{Active Asteroids Citizen Scientist}





\begin{abstract}
%
We announce the discovery of a main-belt comet (MBC), \objnameFull{}. MBCs are a rare type of main-belt asteroid that display comet-like activity, such as tails or comae, caused by sublimation. Consequently, MBCs help us map the location of solar system volatiles, providing insight into the origins of material prerequisite for life as we know it. However, MBCs have proven elusive, with fewer than 20 found among the 1.1 million known main-belt asteroids. This finding derives from Active Asteroids, a NASA Partner Citizen Science program we designed to identify more of these important objects. After volunteers classified an image of \objname{} as showing activity, we carried out a follow-up investigation which revealed evidence of activity from two epochs spanning nearly a decade. This discovery is timely, with \objname{} inbound towards its 2024 March perihelion passage, with potential activity onset as early as late 2023.
\end{abstract}


\section{Introduction} \label{sec:intro}

Main-belt comets (MBCs) are rare, with fewer than 20 found among the 1.1 million known main-belt asteroids. MBCs represent a subpopulation of the active asteroids, which are small solar system bodies that exhibit comet-like activity (i.e., tails, comae) but have asteroidal orbits \citep{jewittActiveAsteroids2015}. The MBCs are active asteroids that are specifically found in the main asteroid belt, and whose activity is attributed to sublimation \citep{hsiehMainbeltCometsPanSTARRS12015}. Knowledge of these objects and their composition help us map the location of solar system volatiles, thereby improving understanding of the origins of the ingredients for life as we know it.

\section{Methods}
\label{sec:methods}

To find more of these remarkable objects we created the Citizen Science program \textit{Active Asteroids}\footnote{\url{http://activeasteroids.net}}, a NASA Partner. Participants classify images of known minor planets, which we extracted from the Dark Energy Camera (DECam) public archive \citep{chandlerSAFARISearchingAsteroids2018,chandlerSixYearsSustained2019,chandlerCometaryActivityDiscovered2020b,chandlerRecurrentActivityActive2021,chandlerMigratoryOutburstingQuasiHilda2022}, as either active or inactive. We investigate activity candidates by conducting archival image searches and follow-up telescope observations, then report our confirmed discoveries \citep[e.g.,][]{oldroydCometlikeActivityDiscovered2023,chandlerNewActiveAsteroid2023,chandlerDiscoveryDustEmission2023}.

\section{Results}
\label{sec:results}

\begin{figure}[h]
    \centering
    \begin{tabular}{ccc}
    	\includegraphics[width=0.32\linewidth]{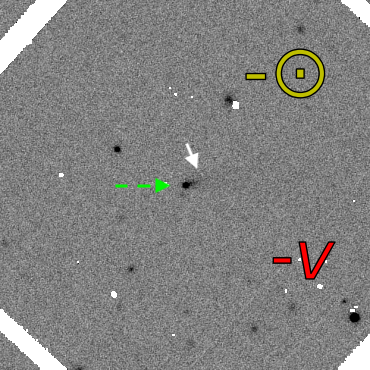} & \includegraphics[width=0.32\linewidth]{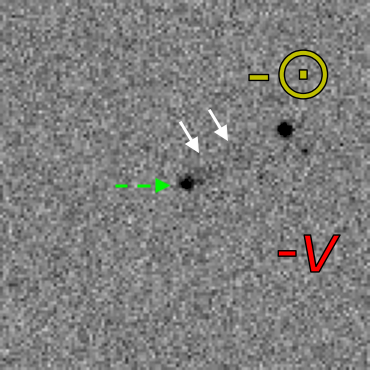} & \includegraphics[width=0.32\linewidth]{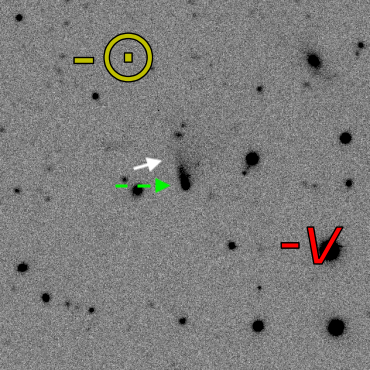}\\
    \end{tabular}
    \caption{\objname\ (green dashed arrows) with a tail (white arrows) oriented on sky roughly towards the anti-solar (yellow -$\odot$) direction, and counter-clockwise from the anti-motion (red $-v$) vector. The FOV of each image is $126'' \times 126''$. 
    \textbf{Left:} UT 2020 September 27 45~s $i$-band GigaPixel1 exposure on the 1.8~m Pan-STARRS~1 (Haleakala). 
    \textbf{Center:} UT 2019 August 31 Zwicky Transient Facility (ZTF) camera on the 48'' Samuel Oschin telescope (Mt. Palomar) 30~s $r$ band exposure.  
    \textbf{Right:} UT 2019 September 30 90~s Dark Energy Camera (DECam) on the 4~m Blanco telescope (Cerro Tololo Inter-American Observatory, Chile); Prop. ID 2019B-1014, PI Olivares, observers F. Olivares, I. Sanchez. This was the image classified as active by \textit{Active Asteroids} volunteers.
    }
    \label{fig:activity}
\end{figure}

One DECam image (Figure \ref{fig:activity}) of \objname{} ($a=2.74$~au, eccentricity $e=0.35$, inclination $i=10.9^\circ$, perihelion distance $q=1.77$~au, aphelion distance $Q=3.72$~au, Tisserand's parameter with respect to Jupiter $T_\mathrm{J}=3.23$, retrieved UT 2023 March 11 from JPL Horizons; \citealt{giorginiJPLOnLineSolar1996}) originally acquired UT 2019 September 30, was unanimously classified as showing activity by \textit{Active Asteroids} volunteers. Our archival investigation revealed additional images (examples provided in Figure \ref{fig:activity}) unambiguously showing activity from two separate orbital epochs: $\sim$10 images from UT 2010 September 27 -- October 10; (true anomaly range of $20.5^\circ<\nu<27.6^\circ$) and $>10$ between UT 2019 August 10 -- November 3 ($-14.2^\circ<\nu<26.5^\circ$). All images of activity were taken when \objname{} was approximately near perihelion passage ($\nu=0^\circ$). When considered with the recurrent activity, this indicates that sublimation is the probable underlying activity mechanism. Hence, given its main-belt orbit, \objname{} is an MBC.

\objname{} is currently observable (as of 2023 March 15; $\nu\sim240^{\circ}$) and currently inbound to its 2024 March 26 perihelion passage, and may become active again as early as 2023 October when it reaches $\nu=290^\circ$, the earliest activity onset point observed to date for an MBC \citep{hsiehObservationalCharacterizationMainBelt2023}, just before its current observing window ends.  It is also important to observe the target prior to its possible reactivation in order to measure properties of the nucleus (e.g., color, absolute magnitude, rotation rate) in the absence of activity, which will impede these measurements. After 2023 October, the object will next be observable from 2024 July to 2025 May ($45^{\circ}\lesssim\nu\lesssim130^{\circ}$).  Observations of the object during these available windows to characterize its expected activity are highly encouraged.

\clearpage
\section*{Acknowledgements}
\begin{acknowledgments}

\textbf{General:} We thank Dr.\ Mark Jesus Mendoza Magbanua (University of California San Francisco) for his ongoing, timely feedback on the project and observing accompaniment.

\textbf{Citizen Science:} We thank Elizabeth Baeten (Belgium) for moderating the Active Asteroids forums. We thank our NASA Citizen Scientists that examined \objname{}: 
Al Lamperti (Royersford, USA), 
Angela Hoffa (Greenfield, USA), 
Carl L. King (Ithaca, USA), 
Jayanta Ghosh (Purulia, India), 
Konstantinos Dimitrios Danalis (Athens, Greece), 
Lydia Yvette Solis (Nuevo, USA), 
Michele T. Mazzucato (Florence, Italy), 
Milton K. D. Bosch MD (Napa, USA), 
Panagiotis J. Ntais (Philothei, Greece), 
and 
Virgilio Gonano (Udine, Italy). 
%
We also thank super-classifiers 
Ivan A. Terentev (Petrozavodsk, Russia) 
and 
Marvin W. Huddleston (Mesquite, USA)
. 
Many thanks to Cliff Johnson (Zooniverse) and Marc Kuchner (NASA) for their ongoing Citizen Science guidance.

\textbf{Funding:} This material is based upon work supported by the NSF Graduate Research Fellowship Program under grant No.\ 2018258765 and grant No.\ 2020303693. 
C.O.C., H.H.H., and C.A.T.\ acknowledge support from NASA grant 80NSSC19K0869. 
W.J.O. and C.A.T.\ acknowledge support from NASA grant 80NSSC21K0114. 
This work was supported in part by NSF awards 1950901 (NAU REU program in astronomy and planetary science). 
Computational analyses were run on Northern Arizona University's Monsoon computing cluster, funded by Arizona's Technology and Research Initiative Fund.

\textbf{Software \& Services:} 
World Coordinate System corrections facilitated by \textit{Astrometry.net} \citep{langAstrometryNetBlind2010}. 
This research has made use of 
NASA's Astrophysics Data System, 
the NASA/IPAC Infrared Science Archive, 
the Institut de M\'ecanique C\'eleste et de Calcul des \'Eph\'em\'erides SkyBoT Virtual Observatory tool \citep{berthierSkyBoTNewVO2006}, 
and 
data and/or services provided by the International Astronomical Union's Minor Planet Center, 
SAOImageDS9, developed by Smithsonian Astrophysical Observatory \citep{joyeNewFeaturesSAOImage2006}. 

\textbf{Facilities \& Instrumentation:} This project used data obtained with the Dark Energy Camera (DECam), which was constructed by the Dark Energy Survey (DES) collaboration. 
This research uses services or data provided by the Astro Data Archive at NSF's NOIRLab. 
Based on observations at Cerro Tololo Inter-American Observatory, NSF’s NOIRLab (NOIRLab Prop. ID 2019B-1014; PI: F. Olivares), 
the Pan-STARRS1 Surveys (PS1) and the PS1 public science archive \citep{chambersPanSTARRS1Surveys2016},
the Zwicky Transient Facility \citep{bellmZwickyTransientFacility2019},
and the CADC Solar System Object Information Search \citep{gwynSSOSMovingObjectImage2012}.
\end{acknowledgments}

\vspace{5mm}
\facilities{
CTIO:4m (DECam), 
IRSA\footnote{\url{https://www.ipac.caltech.edu/doi/irsa/10.26131/IRSA539}}, 
PO:1.2 m (PTF, ZTF), 
PS1
}


\software{
        astropy \citep{robitailleAstropyCommunityPython2013}, 
        {\tt Matplotlib} \citep{hunterMatplotlib2DGraphics2007},
        {\tt NumPy} \citep{harrisArrayProgrammingNumPy2020},
        {\tt pandas} \citep{rebackPandasdevPandasPandas2022}, 
        {\tt SAOImageDS9} \citep{joyeNewFeaturesSAOImage2006},
        {\tt SciPy} \citep{virtanenSciPyFundamentalAlgorithms2020}
          }

\clearpage
\bibliography{zotero}{}

\begin{thebibliography}{}
\expandafter\ifx\csname natexlab\endcsname\relax\def\natexlab#1{#1}\fi
\providecommand{\url}[1]{\href{#1}{#1}}
\providecommand{\dodoi}[1]{doi:~\href{http://doi.org/#1}{\nolinkurl{#1}}}
\providecommand{\doeprint}[1]{\href{http://ascl.net/#1}{\nolinkurl{http://ascl.net/#1}}}
\providecommand{\doarXiv}[1]{\href{https://arxiv.org/abs/#1}{\nolinkurl{https://arxiv.org/abs/#1}}}

\bibitem[{Bellm {et~al.}(2019)Bellm, Kulkarni, Graham, Dekany, Smith, Riddle,
  Masci, Helou, Prince, Adams, Barbarino, Barlow, Bauer, Beck, Belicki, Biswas,
  Blagorodnova, Bodewits, Bolin, Brinnel, Brooke, Bue, Bulla, Burruss, Cenko,
  Chang, Connolly, Coughlin, Cromer, Cunningham, De, Delacroix, Desai, Duev,
  Eadie, Farnham, Feeney, Feindt, Flynn, Franckowiak, Frederick, Fremling,
  {Gal-Yam}, Gezari, Giomi, Goldstein, Golkhou, Goobar, Groom, Hacopians, Hale,
  Henning, Ho, Hover, Howell, Hung, Huppenkothen, Imel, Ip, Ivezi{\'c},
  Jackson, Jones, Juric, Kasliwal, Kaspi, Kaye, Kelley, Kowalski, Kramer,
  Kupfer, Landry, Laher, Lee, Lin, Lin, Lunnan, Giomi, Mahabal, Mao, Miller,
  Monkewitz, Murphy, Ngeow, Nordin, Nugent, Ofek, Patterson, Penprase, Porter,
  Rauch, Rebbapragada, Reiley, Rigault, Rodriguez, {van Roestel}, Rusholme,
  {van Santen}, Schulze, Shupe, Singer, Soumagnac, Stein, Surace, Sollerman,
  Szkody, Taddia, Terek, Van~Sistine, {van Velzen}, Vestrand, Walters, Ward,
  Ye, Yu, Yan, \& Zolkower}]{bellmZwickyTransientFacility2019}
Bellm, E.~C., Kulkarni, S.~R., Graham, M.~J., {et~al.} 2019, Publications of
  the Astronomical Society of the Pacific, 131, 018002,
  \dodoi{10.1088/1538-3873/aaecbe}

\bibitem[{Berthier {et~al.}(2006)Berthier, Vachier, Thuillot, Fernique,
  Ochsenbein, Genova, Lainey, Arlot, Gabriel, Arviset, Ponz, \&
  Solano}]{berthierSkyBoTNewVO2006}
Berthier, J., Vachier, F., Thuillot, W., {et~al.} 2006, in Astronomical {{Data
  Analysis Software}} and {{Systems XV ASP Conference Series}}, Vol. 351
  ({Orem, UT}: {Astronomical Society of the Pacific}), 367

\bibitem[{Chambers {et~al.}(2016)Chambers, Magnier, Metcalfe, Flewelling,
  Huber, Waters, Denneau, Draper, Farrow, Finkbeiner, Holmberg, Koppenhoefer,
  Price, Saglia, Schlafly, Smartt, Sweeney, Wainscoat, Burgett, Grav, Heasley,
  Hodapp, Jedicke, Kaiser, Kudritzki, Luppino, Lupton, Monet, Morgan, Onaka,
  Stubbs, Tonry, Banados, Bell, Bender, Bernard, Botticella, Casertano,
  Chastel, Chen, Chen, Cole, Deacon, Frenk, Fitzsimmons, Gezari, Goessl,
  Goggia, Goldman, Grebel, Hambly, Hasinger, Heavens, Heckman, Henderson,
  Henning, Holman, Hopp, Ip, Isani, Keyes, Koekemoer, Kotak, Long, Lucey, Liu,
  Martin, McLean, Morganson, Murphy, {Nieto-Santisteban}, Norberg, Peacock,
  Pier, Postman, Primak, Rae, Rest, Riess, Riffeser, Rix, Roser, Schilbach,
  Schultz, Scolnic, Szalay, Seitz, Shiao, Small, Smith, Soderblom, Taylor,
  Thakar, Thiel, Thilker, Urata, Valenti, Walter, Watters, Werner, White,
  {Wood-Vasey}, \& Wyse}]{chambersPanSTARRS1Surveys2016}
Chambers, K.~C., Magnier, E.~A., Metcalfe, N., {et~al.} 2016, arXiv.org,
  astro-ph.IM, arXiv:1612.05560

\bibitem[{Chandler {et~al.}(2018)Chandler, Curtis, Mommert, Sheppard, \&
  Trujillo}]{chandlerSAFARISearchingAsteroids2018}
Chandler, C.~O., Curtis, A.~M., Mommert, M., Sheppard, S.~S., \& Trujillo,
  C.~A. 2018, Publications of the Astronomical Society of the Pacific, 130,
  114502, \dodoi{10.1088/1538-3873/aad03d}

\bibitem[{Chandler {et~al.}(2019)Chandler, Kueny, Gustafsson, Trujillo,
  Robinson, \& Trilling}]{chandlerSixYearsSustained2019}
Chandler, C.~O., Kueny, J., Gustafsson, A., {et~al.} 2019, The Astrophysical
  Journal Letters, 877, L12, \dodoi{10/gg3qw6}

\bibitem[{Chandler {et~al.}(2020)Chandler, Kueny, Trujillo, Trilling, \&
  Oldroyd}]{chandlerCometaryActivityDiscovered2020b}
Chandler, C.~O., Kueny, J.~K., Trujillo, C.~A., Trilling, D.~E., \& Oldroyd,
  W.~J. 2020, The Astrophysical Journal Letters, 892, L38, \dodoi{10/gg36xz}

\bibitem[{Chandler {et~al.}(2022)Chandler, Oldroyd, \&
  Trujillo}]{chandlerMigratoryOutburstingQuasiHilda2022}
Chandler, C.~O., Oldroyd, W.~J., \& Trujillo, C.~A. 2022, The Astrophysical
  Journal, 937, L2, \dodoi{10.3847/2041-8213/ac897a}

\bibitem[{Chandler {et~al.}(2021)Chandler, Trujillo, \&
  Hsieh}]{chandlerRecurrentActivityActive2021}
Chandler, C.~O., Trujillo, C.~A., \& Hsieh, H.~H. 2021, The Astrophysical
  Journal, 922, L8, \dodoi{10/gnmckw}

\bibitem[{Chandler {et~al.}(2023{\natexlab{a}})Chandler, Oldroyd, Trujillo,
  Burris, Hsieh, Kueny, Mazzucato, Bosch, \&
  {Shaw-Diaz}}]{chandlerNewActiveAsteroid2023}
Chandler, C.~O., Oldroyd, W.~J., Trujillo, C.~A., {et~al.} 2023{\natexlab{a}},
  Research Notes of the American Astronomical Society, 7, 27,
  \dodoi{10.3847/2515-5172/acbbce}

\bibitem[{Chandler {et~al.}(2023{\natexlab{b}})Chandler, Trujillo, Oldroyd,
  Kueny, Burris, Hsieh, Mazzucato, Bosch, \&
  {Shaw-Diaz}}]{chandlerDiscoveryDustEmission2023}
Chandler, C.~O., Trujillo, C.~A., Oldroyd, W.~J., {et~al.} 2023{\natexlab{b}},
  Research Notes of the American Astronomical Society, 7, 22,
  \dodoi{10.3847/2515-5172/acbb69}

\bibitem[{Giorgini {et~al.}(1996)Giorgini, Yeomans, Chamberlin, Chodas,
  Jacobson, Keesey, Lieske, Ostro, Standish, \&
  Wimberly}]{giorginiJPLOnLineSolar1996}
Giorgini, J.~D., Yeomans, D.~K., Chamberlin, A.~B., {et~al.} 1996, American
  Astronomical Society, 28, 25.04

\bibitem[{Gwyn {et~al.}(2012)Gwyn, Hill, \&
  Kavelaars}]{gwynSSOSMovingObjectImage2012}
Gwyn, S. D.~J., Hill, N., \& Kavelaars, J.~J. 2012, Publications of the
  Astronomical Society of the Pacific, 124, 579, \dodoi{10/f34ggj}

\bibitem[{Harris {et~al.}(2020)Harris, Millman, {van der Walt}, Gommers,
  Virtanen, Cournapeau, Wieser, Taylor, Berg, Smith, Kern, Picus, Hoyer, {van
  Kerkwijk}, Brett, Haldane, {del R{\'i}o}, Wiebe, Peterson,
  {G{\'e}rard-Marchant}, Sheppard, Reddy, Weckesser, Abbasi, Gohlke, \&
  Oliphant}]{harrisArrayProgrammingNumPy2020}
Harris, C.~R., Millman, K.~J., {van der Walt}, S.~J., {et~al.} 2020, Nature,
  585, 357, \dodoi{10.1038/s41586-020-2649-2}

\bibitem[{Hsieh {et~al.}(2015)Hsieh, Denneau, Wainscoat, Sch{\"o}rghofer,
  Bolin, Fitzsimmons, Jedicke, Kleyna, Micheli, Vere{\v s}, Kaiser, Chambers,
  Burgett, Flewelling, Hodapp, Magnier, Morgan, Price, Tonry, \&
  Waters}]{hsiehMainbeltCometsPanSTARRS12015}
Hsieh, H.~H., Denneau, L., Wainscoat, R.~J., {et~al.} 2015, Icarus, 248, 289,
  \dodoi{10.1016/j.icarus.2014.10.031}

\bibitem[{Hsieh {et~al.}(2023)Hsieh, Micheli, Kelley, Knight, Moskovitz,
  Pittichova, Sheppard, Thirouin, Trujillo, Wainscoat, Weryk, \&
  Ye}]{hsiehObservationalCharacterizationMainBelt2023}
Hsieh, H.~H., Micheli, M., Kelley, M. S.~P., {et~al.} 2023, Observational
  {{Characterization}} of {{Main-Belt Comet}} and {{Candidate Main-Belt Comet
  Nuclei}}, \dodoi{10.48550/arXiv.2302.11689}

\bibitem[{Hunter(2007)}]{hunterMatplotlib2DGraphics2007}
Hunter, J.~D. 2007, Computing in Science \& Engineering, 9, 90,
  \dodoi{10.1109/MCSE.2007.55}

\bibitem[{Jewitt {et~al.}(2015)Jewitt, Hsieh, \&
  Agarwal}]{jewittActiveAsteroids2015}
Jewitt, D., Hsieh, H., \& Agarwal, J. 2015, in Asteroids {{IV}} ({Tucson,
  Arizona}: {University of Arizona Press}), 221--241

\bibitem[{Joye(2006)}]{joyeNewFeaturesSAOImage2006}
Joye, W.~A. 2006, in Astronomical {{Data Analysis Software}} and {{Systems XV
  ASP Conference Series}}, Vol. 351, 574--

\bibitem[{Lang {et~al.}(2010)Lang, Hogg, Mierle, Blanton, \&
  Roweis}]{langAstrometryNetBlind2010}
Lang, D., Hogg, D.~W., Mierle, K., Blanton, M., \& Roweis, S. 2010,
  Astronomical Journal, 139, 1782, \dodoi{10.1088/0004-6256/139/5/1782}

\bibitem[{Oldroyd {et~al.}(2023)Oldroyd, Chandler, Trujillo, Burris, Kueny,
  Hsieh, Farrell, DeSpain, Mazzucato, Bosch, {Shaw-Diaz}, \&
  Gonano}]{oldroydCometlikeActivityDiscovered2023}
Oldroyd, W.~J., Chandler, C.~O., Trujillo, C.~A., {et~al.} 2023, Research Notes
  of the American Astronomical Society, 7, 42, \dodoi{10.3847/2515-5172/acc17c}

\bibitem[{Reback {et~al.}(2022)Reback, {jbrockmendel}, McKinney, den Bossche,
  Augspurger, Roeschke, Hawkins, Cloud, {gfyoung}, Sinhrks, Hoefler, Klein,
  Petersen, Tratner, She, Ayd, Naveh, Darbyshire, Garcia, Shadrach, Schendel,
  Hayden, Saxton, Gorelli, Li, Zeitlin, Jancauskas, McMaster, W{\"o}rtwein, \&
  Battiston}]{rebackPandasdevPandasPandas2022}
Reback, J., {jbrockmendel}, McKinney, W., {et~al.} 2022, Pandas-Dev/Pandas:
  {{Pandas}} 1.4.2, Zenodo, \dodoi{10.5281/zenodo.6408044}

\bibitem[{Robitaille {et~al.}(2013)Robitaille, Tollerud, Greenfield,
  Droettboom, Bray, Aldcroft, Davis, Ginsburg, {Price-Whelan}, Kerzendorf,
  Conley, Crighton, Barbary, Muna, Ferguson, Grollier, Parikh, Nair,
  G{\"u}nther, Deil, Woillez, Conseil, Kramer, Turner, Singer, Fox, Weaver,
  Zabalza, Edwards, Azalee~Bostroem, Burke, Casey, Crawford, Dencheva, Ely,
  Jenness, Labrie, Lim, Pierfederici, Pontzen, Ptak, Refsdal, Servillat, \&
  Streicher}]{robitailleAstropyCommunityPython2013}
Robitaille, T.~P., Tollerud, E.~J., Greenfield, P., {et~al.} 2013, Astronomy \&
  Astrophysics, 558, A33, \dodoi{10/gfvntd}

\bibitem[{Virtanen {et~al.}(2020)Virtanen, Gommers, Oliphant, Haberland, Reddy,
  Cournapeau, Burovski, Peterson, Weckesser, Bright, {van der Walt}, Brett,
  Wilson, Millman, Mayorov, Nelson, Jones, Kern, Larson, Carey, Polat, Feng,
  Moore, VanderPlas, Laxalde, Perktold, Cimrman, Henriksen, Quintero, Harris,
  Archibald, Ribeiro, Pedregosa, \& {van
  Mulbregt}}]{virtanenSciPyFundamentalAlgorithms2020}
Virtanen, P., Gommers, R., Oliphant, T.~E., {et~al.} 2020, Nature Methods, 17,
  261, \dodoi{10.1038/s41592-019-0686-2}

\end{thebibliography}
\bibliographystyle{aasjournal}



\end{document}